\documentstyle[spie]{article}

\newcommand{\rb}[1]{\raisebox{1.5ex}[-1.5ex]{#1}}

\def\la{\mathrel{\mathchoice {\vcenter{\offinterlineskip\halign{\hfil
$\displaystyle##$\hfil\cr<\cr\sim\cr}}}
{\vcenter{\offinterlineskip\halign{\hfil$\textstyle##$\hfil\cr
<\cr\sim\cr}}}
{\vcenter{\offinterlineskip\halign{\hfil$\scriptstyle##$\hfil\cr
<\cr\sim\cr}}}
{\vcenter{\offinterlineskip\halign{\hfil$\scriptscriptstyle##$\hfil\cr
<\cr\sim\cr}}}}}
\input{psfig}

\title{Computer simulations of interferometric imaging with the\\
       VLT interferometer and its AMBER instrument} 

\author{ F. Przygodda, T. Bl\"ocker, K.-H. Hofmann, and G. Weigelt
\skiplinehalf 
Max-Planck-Institut f\"ur Radioastronomie, Auf dem H\"ugel 69, 53121 Bonn, Germany
}

\begin{document} 
  \maketitle 

\begin{abstract}
We present computer simulations of interferometric imaging with the
Very Large Telescope Interferometer (VLTI)
of the European Southern Observatory (ESO) 
and the Astronomical MultiBEam Recombiner (AMBER) phase-closure instrument.
These simulations include both the
astrophysical modelling of a stellar object by radiative transfer
calculations and the simulation of  light propagation from the object to the
detector (through  atmosphere, telescopes, and the AMBER instrument),
simulation of photon noise and detector read-out noise, and finally data
processing of the interferograms. The results show the dependence of the
visibility error bars on the following observational parameters:
different seeing during the observation of
object and reference star
(Fried parameters $r_{0,{\rm object}}$ and $r_{0,{\rm ref.}}$
ranging between 0.9\,m and 1.2\,m),
different residual tip-tilt error
($\delta_{\rm tt,object}$ and  $\delta_{\rm tt,ref.}$ ranging between
0.1\% and 20\% of the Airy disk diameter),
and object brightness ($K_{\rm object}$=0.7\,mag to 10.2\,mag,
$K_{\rm ref.}$=0.7\,mag).
Exemplarily, we
focus on stars in late stages of stellar evolution and study one of its
key objects, the dusty supergiant IRC\,+10\,420 that is rapidly evolving on
human timescales.
We show computer simulations of VLT interferometry
(visibility and phase closure measurements)
of  IRC\,+10\,420 with
two and three Auxiliary Telescopes
(ATs; AMBER wide-field mode, i.e.\ without fiber optics spatial filters)
and discuss whether the visibility accuracy is
sufficient to distinguish between different theoretical model predictions.
\end{abstract}


\section{INTRODUCTION}
\label{sect:intro}  
The Very Large Telescope Interferometer\cite{GlinEtal2000} (VLTI)
of the European Southern Observatory (ESO)
with its four 8.2\,m unit
telescopes (UTs) and three 1.8\,m auxiliary telescopes (ATs) will certainly
establish
a new era of optical and infrared interferometric imaging within the next few
years. With a maximum baseline of up to more than 200\,m,
the VLTI will allow the study of astrophysical key objects with
unprecendented resolution opening up new vistas to a better understanding of
their physics. 

The near-infrared focal plane instrument of the VLTI,
the Astronomical MultiBEam Recombiner\cite{PetEtal98,PetEtal2000}
(AMBER),
will operate between 1 and 2.5\,$\mu$m and for up to three beams allowing
the measurement of closure phases. In a second stage
its wavelength coverage is planned to be extended to 0.6\,$\mu$m.
Objects as faint as  $K=20$\,mag are expected to be observable with
AMBER when a bright reference star is available, and as faint as
$K=14$\,mag otherwise.

Among the astrophysical key issues\cite{RichEtal2000}
are, for instance,
young stellar objects, active galactic nuclei and stars in late
stages of stellar evolution. 
The simulation of interferometric imaging of a stellar object
consists in principle of two components: \\[0.5ex]
\begin{minipage}[t]{5ex}
\hspace*{1ex} (i)
\end{minipage}
\begin{minipage}[t]{0.95\textwidth}
the calculation of an astrophysical model of the object, typically based on
           radiative transfer calculations predicting,
           e.g., its intensity distribution.
           To obtain a robust and non-ambiguous model,
           it is of particular importance to take 
           {\it diverse} observational constraints into account, for instance 
           the spectral energy distribution and visibilities.
\end{minipage} \\[0.3ex]
\begin{minipage}[t]{5ex}
\hspace*{1ex} (ii)
\end{minipage}
\begin{minipage}[t]{0.95\textwidth}
the determination of the interferometer's response to this intensity signal,
i.e.\ the simulation of light propagation in the atmosphere and the
interferometer.
\end{minipage} \\[1.0ex]
Often, only one of the above parts is considered in full detail.
The aim of this study is to combine both efforts and to present a computer
simulation
of the VLTI performance for observations of one object class, the dusty
supergiants. For this purpose, we calculated a detailed radiative transfer
model for one of its most outstanding representatives, the supergiant
IRC\,+10\,420, and carried out
computer simulations of VLTI visibility measurements.
The goal is to estimate how
accurate visibilities can be measured with the VLTI in this particular
but not untypical case, to discuss if the accuracy is sufficient to distinguish
between different theoretical model predictions and to study on what the
accuracy is dependent.

\section{The supergiant IRC\,+10\,420: Evolution on human timescales}
The star IRC\,+10\,420 (= V\,1302~Aql)
is an outstanding object for the study of stellar evolution.
Its spectral type changed from
F8\,I$_{\rm a}^{+}$ in 1973\cite{HumStrMurLow73} 
to mid-A today\cite{OudGroeMatBloSah96,KloChePan97}
corresponding to an increase of its effective temperature 
of 1000-2000\,K within only 25\,yr.
It is heavily obscured by circumstellar dust due to strong mass loss with
rates typically of the
order of several $10^{-4}$\,M$_{\odot}$/yr (M$_{\odot}$: solar mass)
\cite{KnaMor85,OudGroeMatBloSah96}.
IRC\,+10\,420 is believed to be a massive luminous star
of initially $\sim 20$ to 40 M$_{\odot}$ 
currently being
observed in its rapid transition from the red supergiant stage to the
Wolf-Rayet ddphase\cite{MutEtal79,NedBow92,JonHumGehEtal93,OudGroeMatBloSah96,KloChePan97}. Wolf-Rayet stars, in turn, finally evolve into a supernova
explosion. IRC\,+10\,420 is the {\it only}
object observed until now in its transition phase to the Wolf-Rayet stage.

Several infrared speckle and coronogra\-phic
observations
\cite{DyckEtal84,RidgEtal86,CobFix87,ChrEtal90,KastWein95}
were conducted to study the dust shell of IRC\,+10\,420.
The most recent study
of Bl\"ocker et al.~\cite{BloeckEtal99}
reports the first diffraction-limited  73\,mas
(mas: milli-arcsecond)
bispectrum speckle interferometry of IRC\,+10\,420
and presents the first radiative transfer calculations
that model {\it both} the spectral energy distribution
{\it and} the visibility of this key object.
We will briefly describe
the main results and conclusions of this study which will serve as
astrophysical input for the VLTI computer simulations presented in the
next section. 
\begin{figure}[hbtp]
\begin{center}
  \begin{tabular}{c}
    \psfig{figure=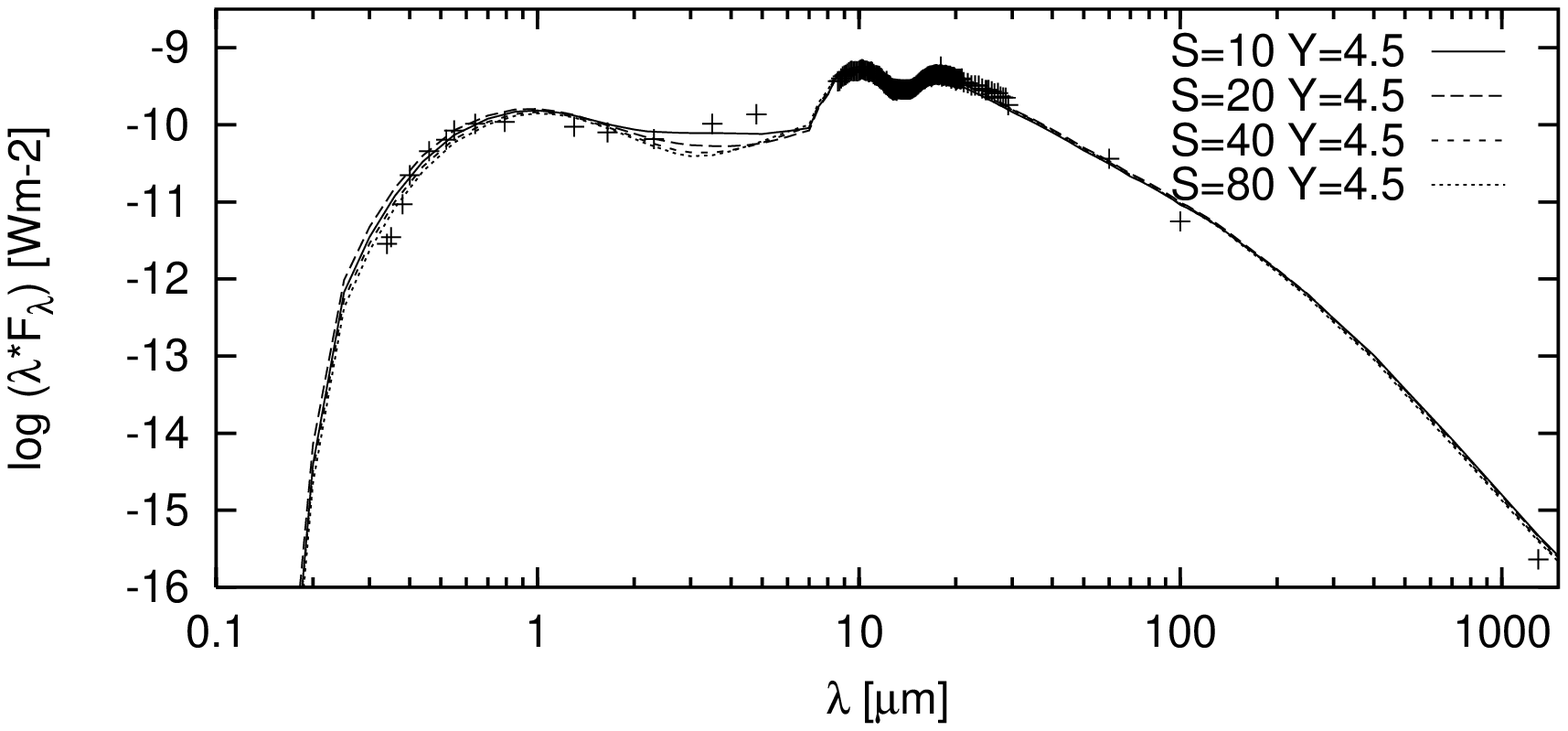,width=0.65\textwidth} \\
    \psfig{figure=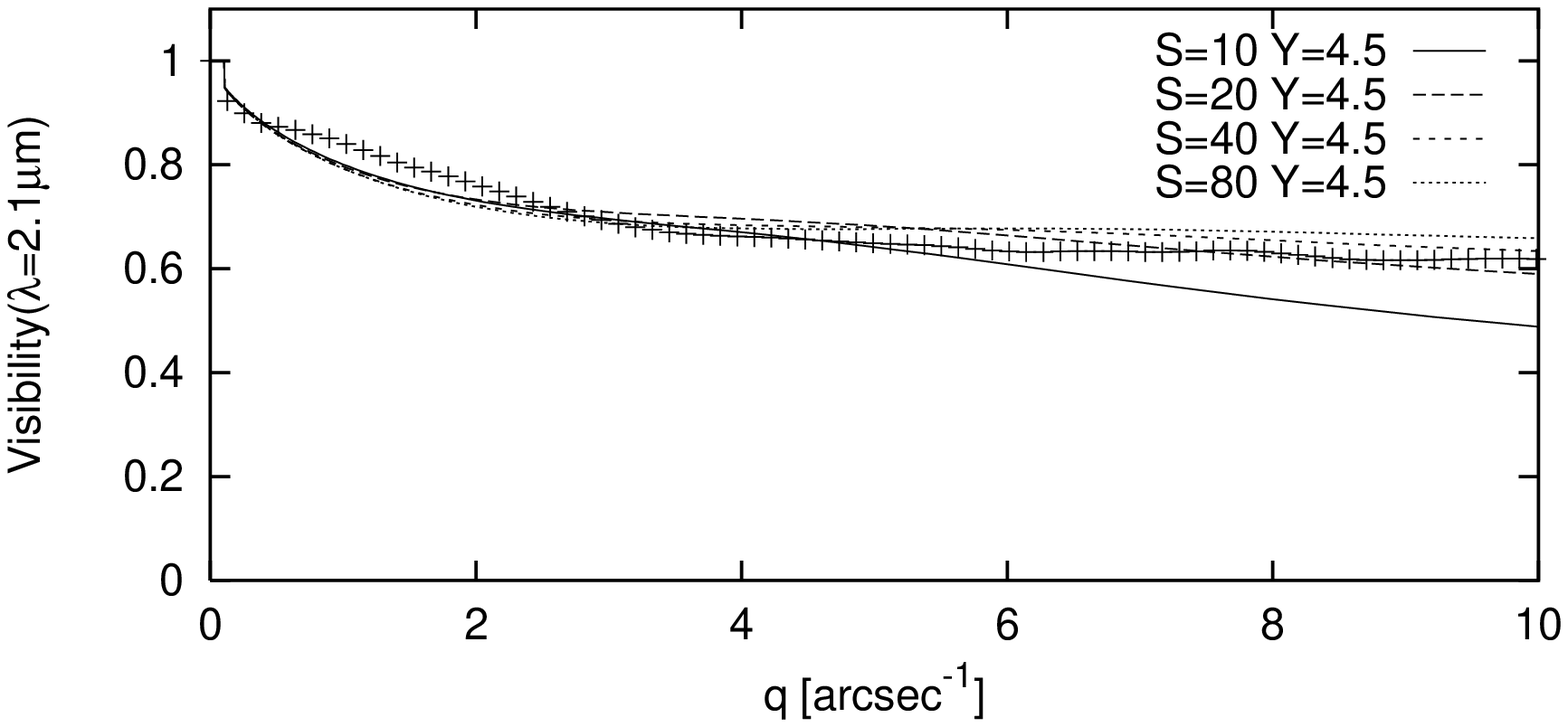,width=0.65\textwidth} \\
\hspace*{-10mm}  \psfig{figure=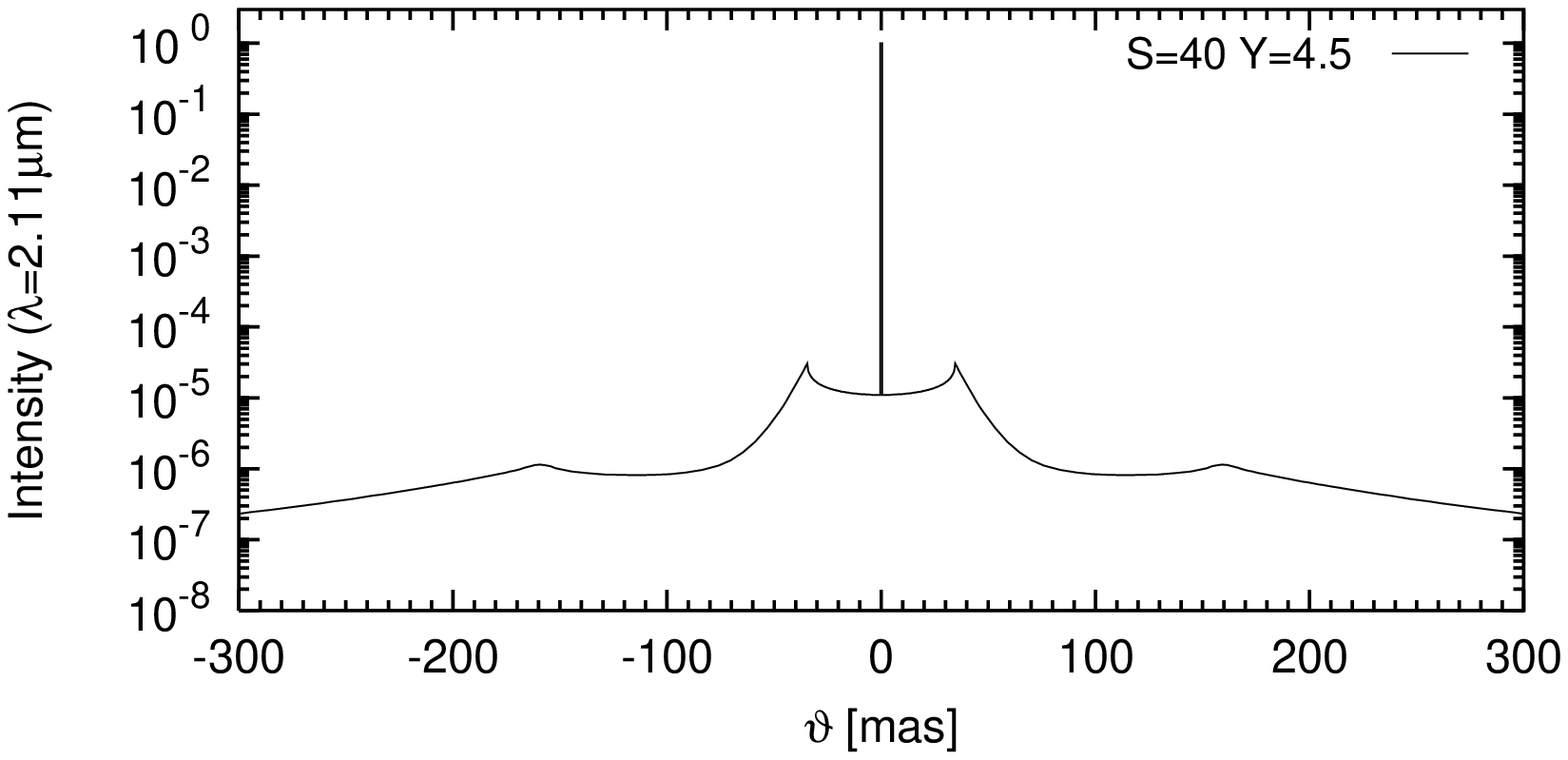,width=0.69\textwidth}  
\end{tabular}
\end{center}
\caption[Fsedvisidens17]
{
SED ({\bf top}) and 2.11\,$\mu$m visibility ({\bf middle})
for superwind models with amplitudes $S$ (=factors of density enhancement)
at $Y=r_{\rm sw}/r_{1}=4.5$, as well as 
normalized intensity vs.\ angular displacement $\vartheta$  ({\bf bottom})
for the best model with  $Y=r_{\rm sw}/r_{1}=4.5$ and $S=40$. In the bottom
panel the (unresolved) central peak belongs to the central star.
The inner hot rim of the circumstellar shell has a radius of 35\,mas, and
a cool component is located at 155\,mas.
Model parameters are:
black body, $T_{\rm eff}=7000$\,K, $T_{1}=1000$\,K,
$\tau_{0.55\mu{\rm m}}=7.0$, silicate dust grains\cite{DraLee84},
standard grain size distribution\cite{MRN77} with
$a_{\rm max}=0.45\,\mu$m,
outer dust-shell boundary=$10^{4} R_{\ast}$ ($R_{\ast}$: stellar radius),
The inner shell region ($Y \le 4.5$) obeys a $1/r^{2}$ density distribution,
the outer shell region  a $1/r^{1.7}$ density distribution.
The symbols
refer to the observations (see text) corrected for interstellar extinction of
$A_{V}=5 {\rm mag}$.
[from Ref.~18]
}                                      \label{Fsedvisidens17}
\end{figure}

Fig.~\ref{Fsedvisidens17} shows the 
SED\cite{OudGroeMatBloSah96,JonHumGehEtal93,WalEtal91,CraEtal76}
and the reconstructed 2.11\,$\mu$m visibility function\cite{BloeckEtal99}
of IRC\,+10\,420. The visibility 0.6 at frequencies $>4$\,cycles/arcsec
indicates, for instance,
that the stellar contribution to the total flux is $\sim$\,60\%
and the dust shell contribution is $\sim$\,40\%.

An extensive grid of radiative transfer models was calculated for the 
dust shell of IRC\,+10\,420
assuming spherical symmetry and
considering black bodies and model atmospheres
as central sources of radiation, different silicates, grain-size and
density distributions,  various dust temperatures at the shell's inner boundary
(determining the radius of the shell's inner boundary) and optical depths.
We refer to Bl\"ocker et al.~\cite{BloeckEtal99}
for a full description of the model grid. 
It turned out that  the observed dust-shell properties cannot be matched by
single-shell models but require multiple components with different density
distributions.
The best model was found  
for a dust shell with a dust temperature of $T_{1}=1000$\,K at its inner radius
of $r_{1}= 69\,R_{\ast}$ ($R_{\ast}$: stellar radius). At a distance of
$r_{\rm sw} = 308\,R_{\ast}$ ($Y = r_{\rm sw}/r_{1} = 4.5$)
the density was enhanced by a factor of $S=40$ and its slope within the shell
changed from $1/r^{2}$ to $1/r^{1.7}$. 
The corresponding fits for SED and  2.11\,$\mu$m visibility are shown in
Fig.~\ref{Fsedvisidens17}.
The  shell's model intensity distribution is shown in
Fig.~\ref{Fsedvisidens17} and was found to be ring-like due to a 
limb-brightened dust-condensation zone.
The ring diameter is equal to the inner diameter of the hot shell
($\sim 69$\,mas), and the diameter of the central star amounts to
$\sim 1$\,mas.

This two-component model  can be interpreted in terms of a termination of an
enhanced mass-loss phase roughly 90 yr ago. 
The assumption that IRC\,+10\,420 had passed through a
superwind phase in its recent history is in line with
its evolutionary status of
an object in transition from the Red-Supergiant to the Wolf-Rayet phase.
%
\section{Interferometry with the VLTI and the AMBER instrument}
In the previous section, the spatial intensity profile of the dusty supergiant
IRC\,+10\,420 (see Fig.~\ref{Fsedvisidens17})
was derived by means of radiative
transfer models and their comparison with photometric and interferometric
observations. This $2.11\,\mu$m intensity profile will serve as object
intensity profile in the simulation of monochromatic VLTI observations. 
The next steps are the simulation of  light propagation from the object to
the detector (through  atmosphere, telescopes, and the AMBER wide-field mode
instrument),
simulation of photon noise and detector read-out noise, and finally data
processing of the interferograms.
\subsection{Computer simulation of interferometric imaging}
\begin{figure}
\begin{center}
  \begin{tabular}{c}
    \psfig{figure= 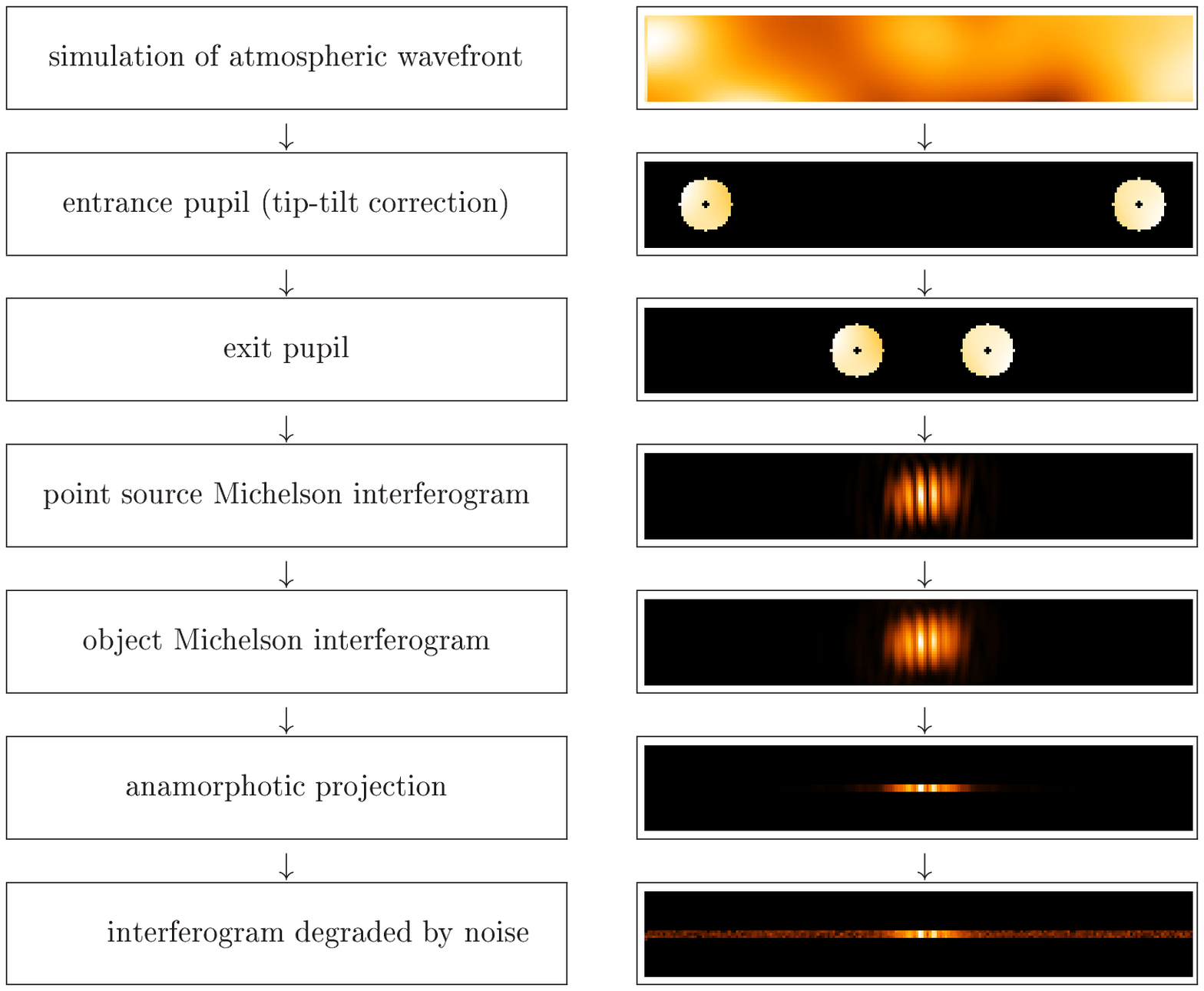,width=1.0\textwidth}
\end{tabular}
\end{center}
\caption[fluss2d]
{
Flow chart of the simulation of VLTI AMBER interferograms
(wide-field mode, i.e. without fiber optics spatial filtering).
} \label{Ffluss2d}
\end{figure}
Fig.~\ref{Ffluss2d} shows a flow chart of our simulation of interferometric
imaging with the VLTI
(ATs, or UTs with adaptive optics)
and the AMBER camera in the wide-field
mode (i.e. without fiber optics spatial filtering).
The  total number of detectable photons $N_{\rm tot}$ is given by the
object brightness, the wavelength and
band-width, the collecting area of the telescope, the total transmission
of the optics, and the quantum efficiency of the detector\cite{MamEtal93}.
Since we assume that the observations are carried out with a spectral
resolution of 
$R=\lambda/\Delta \lambda=70$, the number of photons 
available in one spectral channel for observations
in the $K$-band is given by
$N_{\rm channel} = N_{K,{\rm tot}} \cdot R_{K}/R$,
with $R_{K}= 2.2\,\mu{\rm m}/0.4\,\mu{\rm m}=5.5$ being the 
spectral resolution of the $K$-band
(i.e. $R/R_{K} \sim 13$ is the number of spectral channels within the
$K$-band; $N_{K,{\rm tot}}$ is the total number of $K$-band photons).
The simulations described below
refer to observations in one spectral channel (see also Table~\ref{Tsimu1}).
The signal-to-noise ratio of the reconstruction can be increased by
averaging the reconstructions of all  $R/R_{K}$ channels. 
\begin{table} 
\begin{center}       
\caption{Parameters of VLTI-AT/AMBER simulations:
Wavelength, object brightness, optical
throughput (=throughput of VLTI-AT\,+\,AMBER in the wide-field mode,
AT mirror size: 1.8\,m),
read-out noise and quantum
efficiency of the detector, exposure time, and photon number $N_{\rm channel}$
per frame and per {\it one}
spectral channel for the lowest spectral resolution of AMBER of
$R=\lambda/\Delta \lambda=70$ (instrumental AMBER characteristics taken from 
Malbet et al.\cite{MalEtal2000})
}
\vspace*{1ex}
\label{Tsimu1}
{\footnotesize
\begin{tabular}{|c|c|c|c|c|c|c|}
\hline
\rule[-1ex]{0pt}{3.5ex} wavelength   & $K$ brightness  & opt.\ throughput &
    read-out noise & quant.\ efficiency & exposure & photon number $N_{\rm channel}$ 
\\ \hline 
\rule[-1ex]{0pt}{3.5ex} $2.11\,\mu$m & $0.7 - 10.2$\,mag  & 0.108           &
    15\,e$^{-}$    &  0.6               & 50\,ms   & $ 4.42 \cdot 10^{6}  -
                                                       7.00 \cdot 10^{2}$ 
\\ \hline
\end{tabular}
}
\end{center}
\end{table}

In a first step, an array of
Gaussian distributed random numbers is generated and convolved  with the
correlation function of the atmospheric refraction index variations
(see Roddier\cite{Rod81})
in order to generate wavefronts degraded by atmospheric
turbulence.
The typical size of the atmospheric turbulence cells is given
by the Fried parameter $r_{0}$. After the simulation of the entrance pupil,
the next step incorporates
the tip-tilt correction of the wavefront over each subpupil,
but we allow for a residual tip-tilt  error $\delta_{\rm tt}$.
In the next step a typical Michelson output pupil is simulated
by shifting both subpupils from the entrance pupil position,
with a subpupil separation equal to the baseline, to the output
pupil position where both subpupils are only separated by $\sim$ one
subpupil diameter (pupil reconfiguration).
The output pupil is chosen such that (i) in the optical transfer function
the off-axis peaks are separated from the central peak, and (ii)
the interferograms are sampled with the smallest number of pixels to assure
the lowest influence of detector noise.
The following step includes
light propagation through the beam combiner lens
to the focal plane. The squared modulus of  the
Fourier transform of the complex amplitude in front of the beam combiner lens
yields the intensity distribution of a Michelson interferogram of a point
source.
In the next step the required object intensity distribution is simulated
(given here by the intensity distribution of IRC\,+10\,420)
to obtain the Michelson interferogram of the object:
The Fourier transformation of the object intensity distribution, calculated
at those spatial frequencies covered by the the simulated interferometer
baseline vector, is multiplied with the off-axis peaks of the transfer
function of the generated Michelson interferogram\cite{Tall2}.
Finally, Poisson photon-noise and detector read-out noise is injected to the
interferograms.
The noise level depends, among other parameters, on the
number of detectable photons,
the total optical throughput of the
interferometer (VLTI\,+\,AMBER in the wide-field mode)
and the quantum efficiency of the detector.
Details\cite{MalEtal2000} of the simulation parameters specific to
VLTI-AT/AMBER, as optical throughput, detector read-out noise
and quantum efficiency, are given in Table~\ref{Tsimu1}.

\subsection{Computer simulations of visibility measurements}
We performed simulations of IRC\,+10\,420 visibility observations
in the VLTI-AT/AMBER wide-field mode
and studied the influence of various observational parameters on the
visibility accuracy.
Visibility error bars were obtained for the following
observational parameters:
\begin{enumerate}
\item different seeing during the observation of
      object and reference star
      (Fried parameters $r_{0,{\rm object}}$ and $r_{0,{\rm ref.}}$
      ranging between 0.9\,m and 1.2\,m),
\item different residual tip-tilt error
      ($\delta_{\rm tt,object}$ and  $\delta_{\rm tt,ref.}$ ranging between
      0.1\% and 20\% of the Airy disk diameter),
\item different object brightness ($K_{\rm object}$=0.7\,mag to 10.2\,mag,
       $K_{\rm ref.}$=0.7\,mag).
\end{enumerate}

All computer experiments are based on 2400 interferograms of
the $S=40$ intensity profile (see Fig.~\ref{Fsedvisidens17}) of IRC\,+10\,420
and refer to observations in one of the $R/R_{K} \sim 13$ spectral channels.
The error bars correspond to the reduction of six statistically independent
data sets and refer to the standard deviation $\sigma$.
The dependence of the error bar on the number of data sets has been verified
(see below).
In the computer experiments (1) and (2) mentioned above, 
object and reference star were assumed to have
the same brightness. We chose $K$=0.7\,mag (see Table~\ref{Tsimu1})
in order to minimize brightness effects. 
Experiment (3) simulates IRC\,+10\,420 itself ($K$=3.5\,mag)
and fainter objects, assuming
object and reference star to have the same seeing and
residual tip-tilt error ($r_{0,{\rm object}}$=$r_{0,{\rm ref}}$=0.9\,m;
$\delta_{\rm tt,object}$=$\delta_{\rm tt,ref}$=0.1\%). These simulations
supersede the corresponding results of a first exploratory study
\cite{BloeckerEtal2000} 
due to improved numerics and a broader statistical basis. 

\begin{figure}
\begin{minipage}{0.46\textwidth}
\begin{center}
\begin{tabular}{|c|c|c|c|c|c|} \hline
 &  & $r_0$ [m] & $\delta_{\rm tt} [\%]$ &  $K$ [mag] & $\sigma_{V}$\\
\hline \hline 
       & obj. & 0.9 & 0.1  & 0.7 & \\
\rb{A} & ref. & 0.9 & 0.1  & 0.7 & \rb{0.0067} \\ 
\hline  
       & obj. & {\bf 1.1} & 0.1  & 0.7 & \\
\rb{B} & ref. & {\bf 1.1} & 0.1  & 0.7 & \rb{0.0047} \\ 
\hline 
       & obj. & {\bf 1.2} & 0.1  & 0.7 & \\
\rb{C} & ref. & {\bf 1.2} & 0.1  & 0.7 & \rb{0.0036} \\ 
\hline \hline 
       & obj. & {\bf 1.0} & 0.1  & 0.7 & \\
\rb{D} & ref. &       0.9 & 0.1  & 0.7 & \rb{0.0214} \\ 
\hline 
       & obj. & {\bf 1.1} & 0.1  & 0.7 & \\
\rb{E} & ref. &      0.9  & 0.1  & 0.7 & \rb{0.0376} \\ 
\hline 
       & obj. & {\bf 1.2} & 0.1  & 0.7 & \\
\rb{F} & ref. &       0.9 & 0.1  & 0.7 & \rb{0.0551} \\ 
\hline 
\end{tabular}
\end{center}
\end{minipage}
\begin{minipage}{0.52\textwidth}
 \hspace{1cm} \\[3.0ex]
  \begin{tabular}{c}
    \psfig{figure=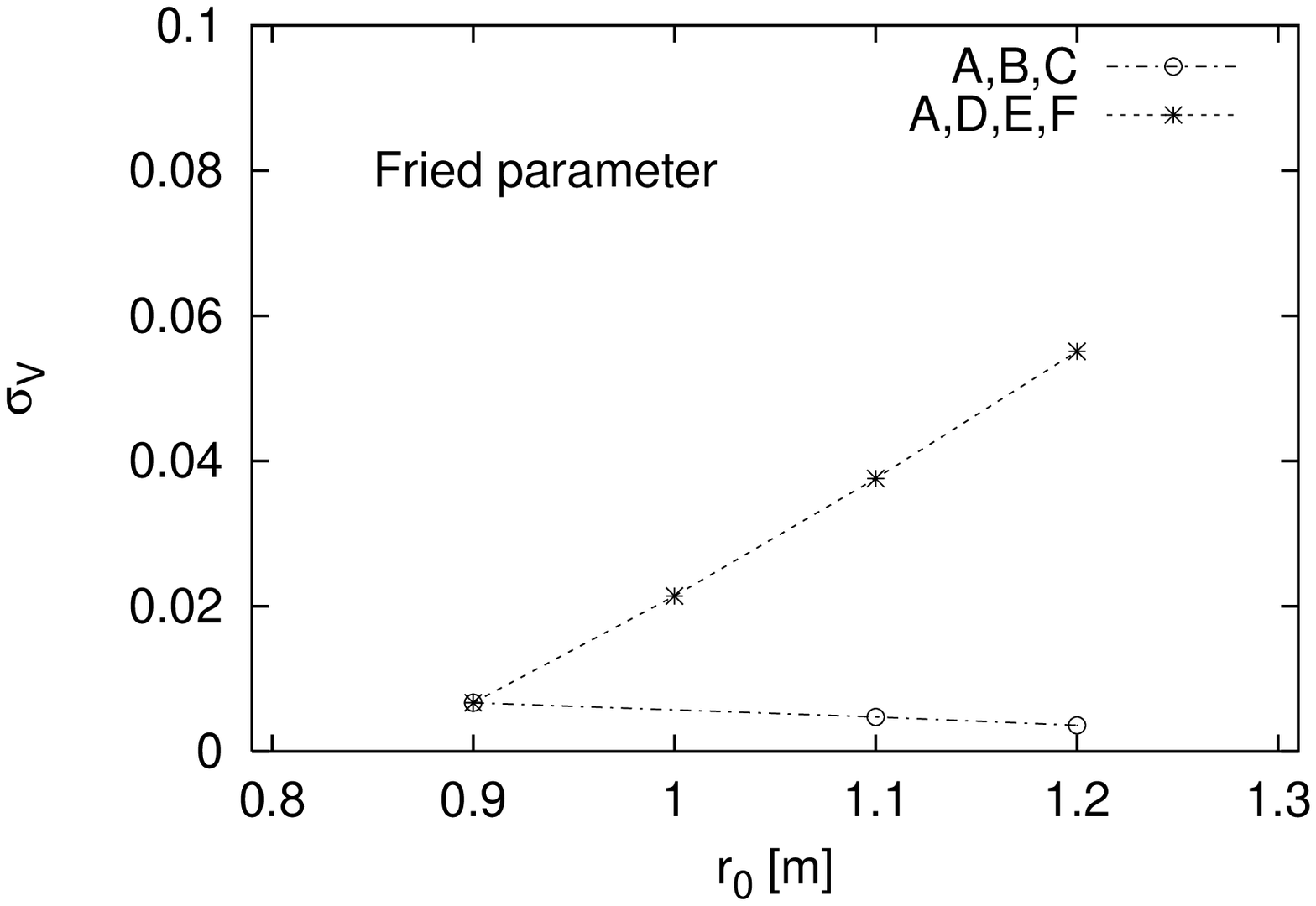,width=1.00\textwidth} \\
  \end{tabular}
\end{minipage}

\begin{minipage}{0.46\textwidth}
\begin{center}
\begin{tabular}{|c|c|c|c|c|c|} \hline
 & &  $r_0$ [m] & $\delta_{\rm tt} [\%]$ &  $K$ [mag] & $\sigma_{V}$\\
\hline \hline 
       & obj. & 0.9 & 0.1  & 0.7 & \\
\rb{A} & ref. & 0.9 & 0.1  & 0.7 & \rb{0.0067} \\ 
\hline  
       & obj. & 0.9 & {\bf 10}  & 0.7 & \\
\rb{G} & ref. & 0.9 & {\bf 10}  & 0.7 & \rb{0.0072} \\ 
\hline 
       & obj. & 0.9 & {\bf 20}  & 0.7 & \\
\rb{H} & ref. & 0.9 & {\bf 20}  & 0.7 & \rb{0.0084} \\ 
\hline \hline 
       & obj. & 0.9 & {\bf 15}  & 0.7 & \\
\rb{I} & ref. & 0.9 &      10   & 0.7 & \rb{0.0080} \\ 
\hline 
       & obj. & 0.9 & {\bf 20}  & 0.7 & \\
\rb{J} & ref. & 0.9 &      10   & 0.7 & \rb{0.0107} \\ 
\hline 
\end{tabular}
\end{center}
\end{minipage}
\begin{minipage}{0.52\textwidth}
 \hspace{1cm} \\[3.0ex]
  \begin{tabular}{c}
    \psfig{figure=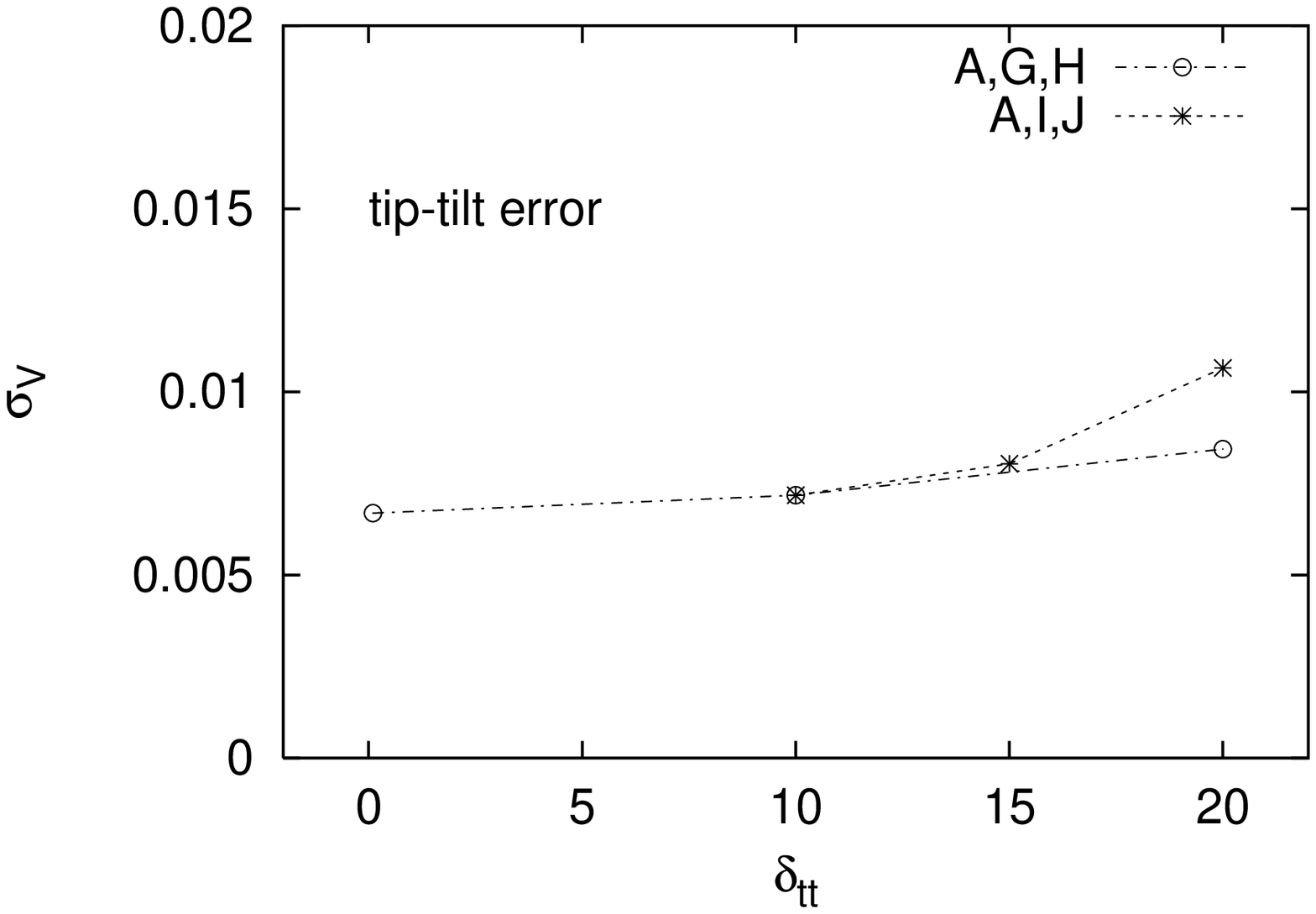,width=1.00\textwidth} \\
  \end{tabular}
\end{minipage}
\begin{minipage}{0.46\textwidth}
\begin{center}
\begin{tabular}{|c|c|c|c|c|c|} \hline 
  &   & $r_0$ [m] & $\delta_{\rm tt} [\%]$ &  $K$ [mag] & $\sigma_{V}$\\ 
\hline \hline 
       & obj. & 0.9 & 0.1  & 0.7 & \\
\rb{A} & ref. & 0.9 & 0.1  & 0.7 & \rb{0.0067} \\ 
\hline  
       & obj. & 0.9 & 0.1  & {\bf 7.2}  & \\
\rb{K} & ref. & 0.9 & 0.1  & 0.7 & \rb{0.0073} \\ 
\hline 
       & obj. & 0.9 & 0.1  & {\bf 8.2}  & \\
\rb{L} & ref. & 0.9 & 0.1  & 0.7 & \rb{0.0123} \\ 
\hline 
       & obj. & 0.9 & 0.1  & {\bf 9.2}   & \\
\rb{M} & ref. & 0.9 & 0.1  & 0.7 & \rb{0.0156} \\ 
\hline 
       & obj. & 0.9 & 0.1  & {\bf 10.2}  & \\
\rb{N} & ref. & 0.9 & 0.1  & 0.7 & \rb{0.0756} \\ 
\hline 
\end{tabular}
\end{center}
\end{minipage}
\begin{minipage}{0.52\textwidth}
 \hspace{1cm} \\[3.0ex]
  \begin{tabular}{c}
    \psfig{figure= 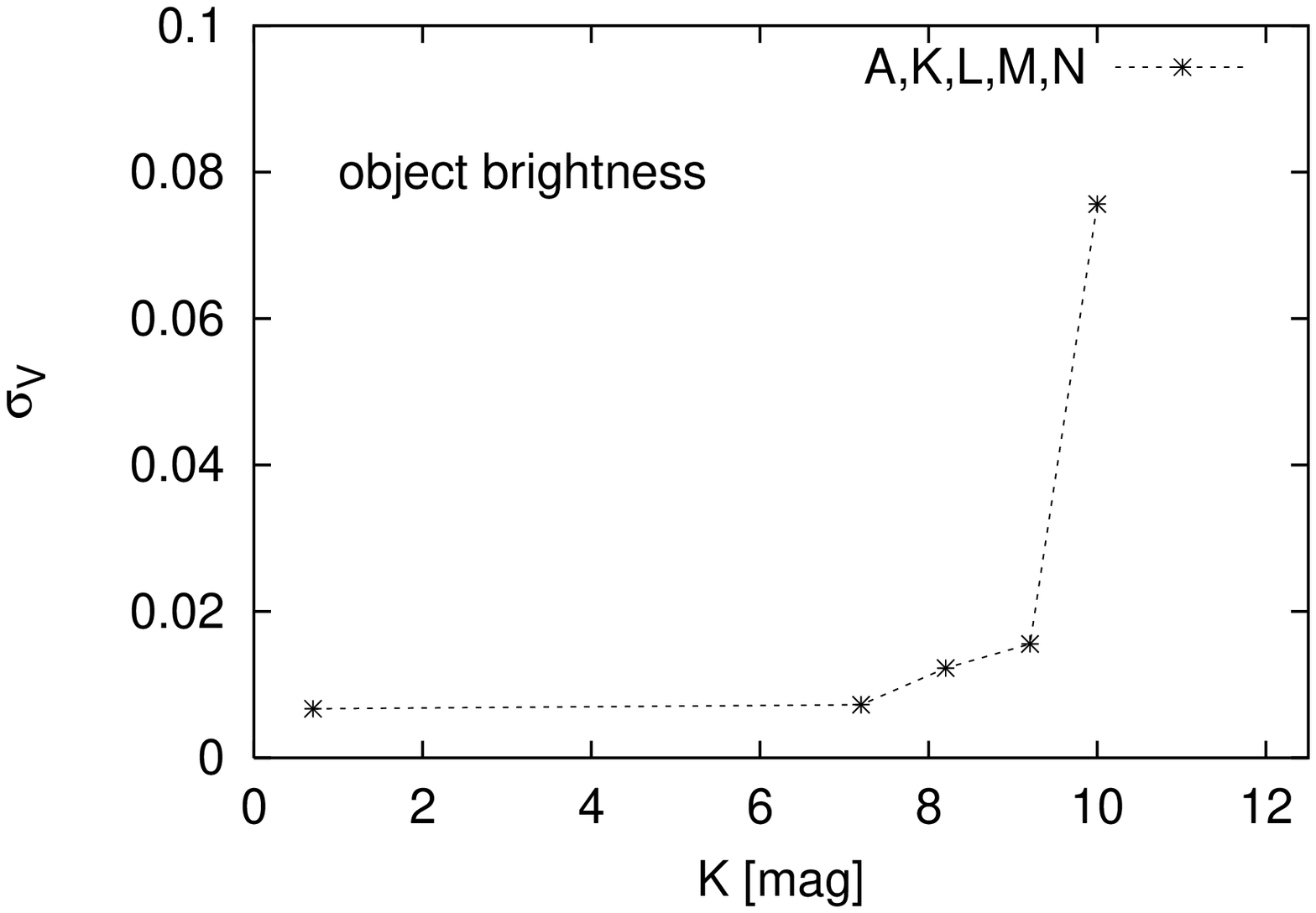,width=1.00\textwidth} \\
  \end{tabular}
\end{minipage}
\caption{\footnotesize
Dependence of the error bars of simulated AT-VLTI/AMBER 
(wide-field mode) observations of IRC\,+10\,420 at 2.11\,$\mu$m on
the {\it seeing conditions} [{\bf top}],
{\it residual tip-tilt error} [{\bf middle}], and
{\it object brightness} [{\bf bottom}]. 
The table columns refer to the Fried paramater $r_{0}$, the residual tip-tilt
error $\delta_{\rm tt}$, the $K$-magnitude and the visibility error
$\sigma_{V}$
(based on  6 statistically independent repititions of each simulation). 
Each AT-VLTI/AMBER simulation refer to $N$=2400 interferograms
of the $S$=40 intensity profile
(see Fig.~\ref{Fsedvisidens17}).
}
\label{Fsimulall}
\end{figure}

The first computer experiment (different seeing conditions) consists of
the simulations A to F (see top panel of Fig.~\ref{Fsimulall}). 
Simulation A 
represents the  case of typical seeing conditions
($r_{0,{\rm object}}$=0.9\,m, $r_{0,{\rm ref.}}$=0.9\,m)
and an almost perfect tip-tilt correction (residual tip-tilt error
$\delta_{\rm tt}=0.1$\%).
It will serve as a reference for the following simulations.
The brightness of object and reference star was chosen to be $K$=0.7\,mag.
The visibility error $\sigma_{V}$ amounts only to $\pm 0.0067$.
The remaining statistical uncertainty due to the limited number of data sets
has been checked by a test calculation taking into account twice as many
statistically independent data sets, i.e. 12 data sets instead of 6.
The visibility error changed by 24\% from $\pm 0.0067$ to
$\pm 0.0051$. However, the general dependencies of the visibility accuracy
on observational parameters (seeing, residual tip-tilt error, brightness)
are only scarcely affected. 

If one considers a better seeing for {\it both} object {\it and}
reference star (A-C),
$\sigma_{V}$ decreases in almost linear proportion with
increasing diameter of the atmospheric turbulence cells (Fig~\ref{Fsimulall}).
However, improved but {\it different} seeing conditions (D-F)
lead to larger visibility errors.
This emphasizes that
equal seeing conditions for object and reference star are more crucial than
excellent seeing conditions of, e.g., the object only in order to obtain
accurate visibilities.

The second set of simulations (A, G-J) illustrates the influence of a
larger residual object tip-tilt error (see middle panel of
Fig.~\ref{Fsimulall}). Increasing
$\delta_{\rm tt}$ from 0.1\% to 20\% for both object and reference star
increases the visibility error from 0.0067 to 0.0084 (A,F,G). 
Like in the simulations of different seeing conditions, it is
essential to have the same tip-tilt error for both object and reference star
in order to achieve high accuracy.
Differences in $\delta_{\rm tt}$ enhance the visibility error
(H,I). 

Finally, in the third experiment of this series 
an IRC\,+10\,420-like intensity distribution is assumed but
much much fainter (i.e. fainter than $K$=3.5\,mag) objects are considered
(see bottom panel of Fig.~\ref{Fsimulall}, simulations A, K-N).
The visibility error stays almost constant up to an $K$ magnitude of 7\,mag.
If the simulated object $K$-magnitude is 9.2\,mag,  
the visibility error has increased by a factor of $\sim 2$,
i.e.\ to $\sigma_{V}=0.0156$, 
but is still acceptable although the photon number decreased to
$N_{\rm channel}=1.76 \cdot 10^{3}$ (i.e.\ by a factor of 2500, see  Table~\ref{Tsimu1}).
Fig.~\ref{Fsimulall} demonstrates that for even fainter objects
noise becomes more and more important leading to an steep increase of the 
visibility error.
However, the signal-to-noise ratio can be improved
($\sim \sqrt{R/R_{K}} \sim 3.6$)
if the visibility
measurements of all  $R/R_{K}$  spectral channels within the $K$-band are
averaged. 
%

%
%
%
\begin{figure}
  \begin{tabular}{c}
    \psfig{figure=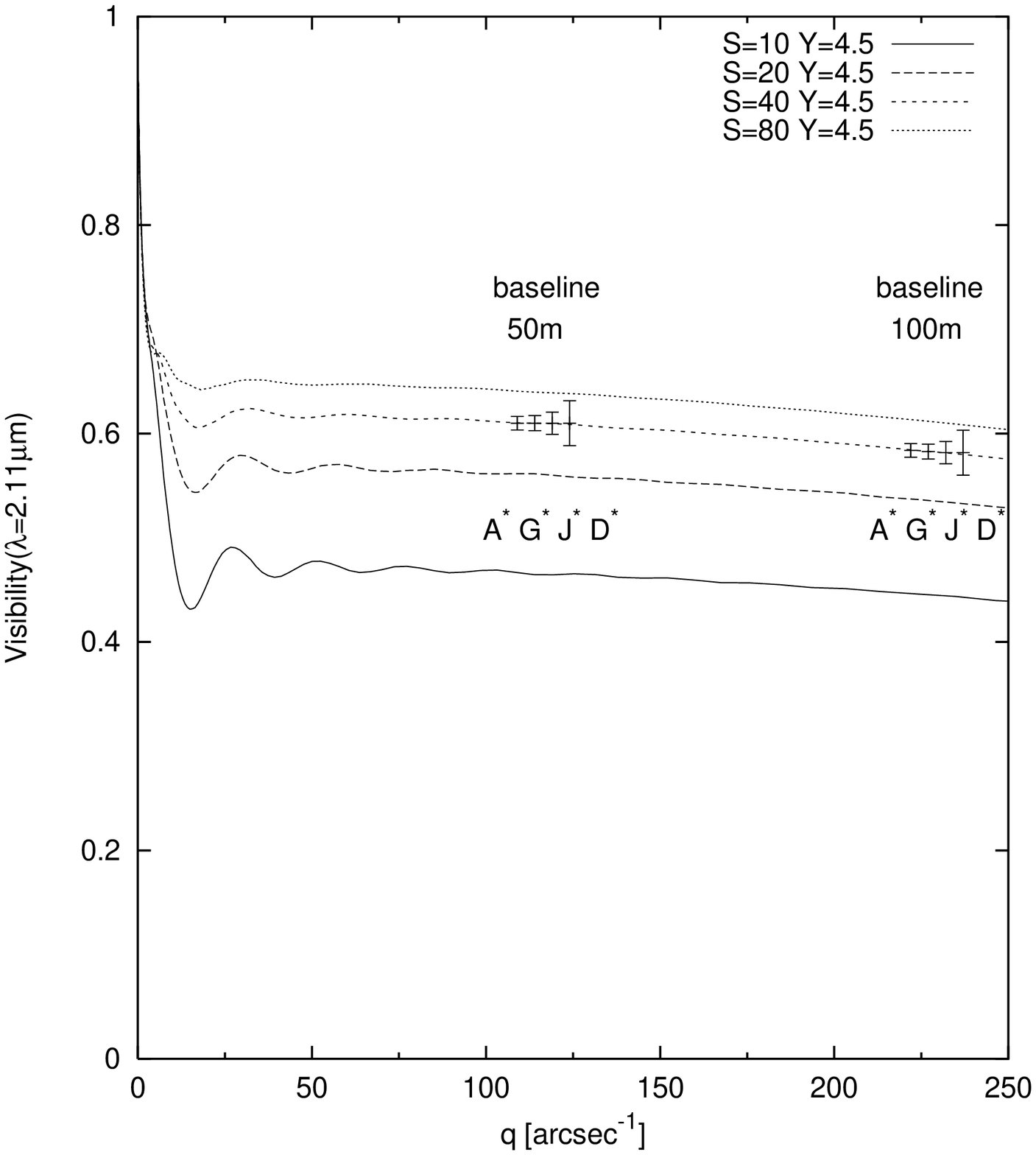,width=0.94\textwidth} \\
  \end{tabular}
\vspace*{-5mm}
\caption[simvisATH]
{
Dependence of the error bars of simulated AT-VLTI/AMBER 
(wide-field mode) observations of IRC\,+10\,420 at 2.11\,$\mu$m on
(i) seeing  differences of object and reference star
observations (Fried parameter, $r_{0}$, differences; simulation D), and
(ii) different residual tip-tilt errors, $\delta_{\rm tt}$,
of object and reference star observations (simulations G and J). 
Lines refer to radiative transfer models of
different superwind amplitudes $S$. The symbols refer to 
AT-VLTI/AMBER simulations (each with $N$=2400 interferograms)
of the $S$=40 intensity profile
(see Fig.~\ref{Fsedvisidens17}).
for baselines of 50 and 100\,m
(100\,m: $q$=227 arc\,sec$^{-1}$).
To better distinguish between the simulations,
the data points belonging to one baseline are somewhat shifted 
with respect to the spatial frequency. 
The error bars are based on  6 statistically independent repetitions of
each simulation. 
The table gives the parameters of the simulations
A$^{\ast}$, G$^{\ast}$, J$^{\ast}$, and D$^{\ast}$ 
(cf.\ Fig.~\ref{Fsimulall}). The asterisks indicate that the
object brightness is here 3.5\,mag as appropriate for IRC\,+10\,420 whereas
the simulations shown in Fig.~\ref{Fsimulall} refer to an object brighness of
0.7\,mag.
} \label{FsimvisATH}
\hspace*{0.7cm}
\begin{minipage}{1.0\textwidth}
\vspace*{-19cm}
\begin{center}      
\begin{tabular}{|c|cc|cc|cc|cc|}
\hline
\rule[-1ex]{0pt}{3.5ex}  &
                        \multicolumn{2}{c}{A$^{\ast}$} &
                        \multicolumn{2}{c}{G$^{\ast}$} & 
                        \multicolumn{2}{c}{J$^{\ast}$} & 
                        \multicolumn{2}{c|}{D$^{\ast}$} 
\\  
\rule[-1ex]{0pt}{3.5ex}  &
                        \multicolumn{2}{c}{base experiment} &
                        \multicolumn{2}{c}{$\delta_{\rm tt}$ dependence} & 
                        \multicolumn{2}{c}{$\delta_{\rm tt}$ dependence} & 
                        \multicolumn{2}{c|}{$r_{0}$ dependence} 
\\  
\rule[-1ex]{0pt}{3.5ex}  &
                         obj & ref & 
                         obj & ref & 
                         obj & ref & 
                         obj & ref  
\\ \hline \hline
\rule[-1ex]{0pt}{3.5ex} $K$-mag. &
                         3.5  &  0.7  &
                         3.5  &  0.7  &
                         3.5  &  0.7  &
                         3.5  &  0.7  
\\ \hline
\rule[-1ex]{0pt}{3.5ex} $r_{0}$ &
                         0.9  &  0.9  &
                         0.9  &  0.9  &
                         0.9  &  0.9  &
                   {\bf 1.0}  &  {\bf 0.9}  
\\ \hline
\rule[-1ex]{0pt}{3.5ex} $\delta_{\rm tt}$ &
                       0.1\%   & 0.1\%   & 
                   {\bf 10\%} & {\bf 10\%}   & 
                   {\bf 20\%} & {\bf 10\%}   & 
                       0.1\%   & 0.1\%    
\\ \hline
\rule[-1ex]{0pt}{3.5ex} $N$ &
                        2400   & 2400   &
                        2400   & 2400   &
                        2400   & 2400   &
                        2400   & 2400   
\\ \hline
\rule[-1ex]{0pt}{3.5ex} $\sigma_{V}$ &
                        \multicolumn{2}{c|}{$ 0.0066$} &
                        \multicolumn{2}{c|}{$ 0.0072$} &
                        \multicolumn{2}{c|}{$ 0.0106$} &
                        \multicolumn{2}{c|}{$ 0.0215$} 
\\ \hline
\end{tabular}
\end{center}
\end{minipage}
\end{figure}
%
%

%
For illustration, Fig.~\ref{FsimvisATH} shows the results of the
simulations A$^{\ast}$, G$^{\ast}$, J$^{\ast}$ and D$^{\ast}$
for baselines of 50 and 100\,m together with the model predictions for
IRC\,+10\,420. The asterisk indicates that within these simulations the
object brightness is that of IRC\,+10\,420, viz.\ $K$=3.5\,mag
(photon number: $3.47  \cdot 10^{5}$)  As it is obvious
from Fig.~\ref{Fsimulall} (bottom panel), this brightness is well within the
range where the visibility error is not very sensitive to brightness changes.
Accordingly, the results are very close to those of the simulations A, G, J,
and D. Again, different seeing conditions (simulation D$^{\ast}$)
for object and reference star
are more crucial than, e.g., residual tip-tilt errors
(simulations G$^{\ast}$ and J$^{\ast}$).
For typical seeing conditions (Fried parameter differences $\la$ 10\%),
the visibility accuracy achievable with the VLTI using the
ATs in the wide-field mode is clearly sufficient to distinguish between
different radiative transfer models of IRC\,+10\,420 and, thus, to prove
(or disprove) theoretical predictions. 

\subsection {Computer simulations of phase-closure imaging}
In addition to the above visibility studies, image reconstruction
simulations were performed as well. For this purpose, modulus (=visibility)
{\it and} phase of the Fourier transform of the object's intensity
distribution were determined utilizing the phase-closure
method\cite{Jen1959}. The phase closure method requires the simultaneous
measurement of the object with at least three telescopes. 
For a corresponding computer experiment we chose three telescopes covering
five configurations. The baseline between telescope 1 and telescope 2 was
8\,m whereas the baseline between  telescope 2 and telescope 3 
amounted to  8, 16, 24, 32 and 40\,m in order to facilitate
a simple recursive algorithm for phase reconstruction from the measured
closure phases.
A non-redundant arrangement of the output pupil
ensures a sufficient separation of the object information in frequency space.

\begin{figure}
\begin{minipage}{0.48\textwidth}
 \hspace{1cm} \\[3.0ex]
  \begin{tabular}{c}
    \psfig{figure=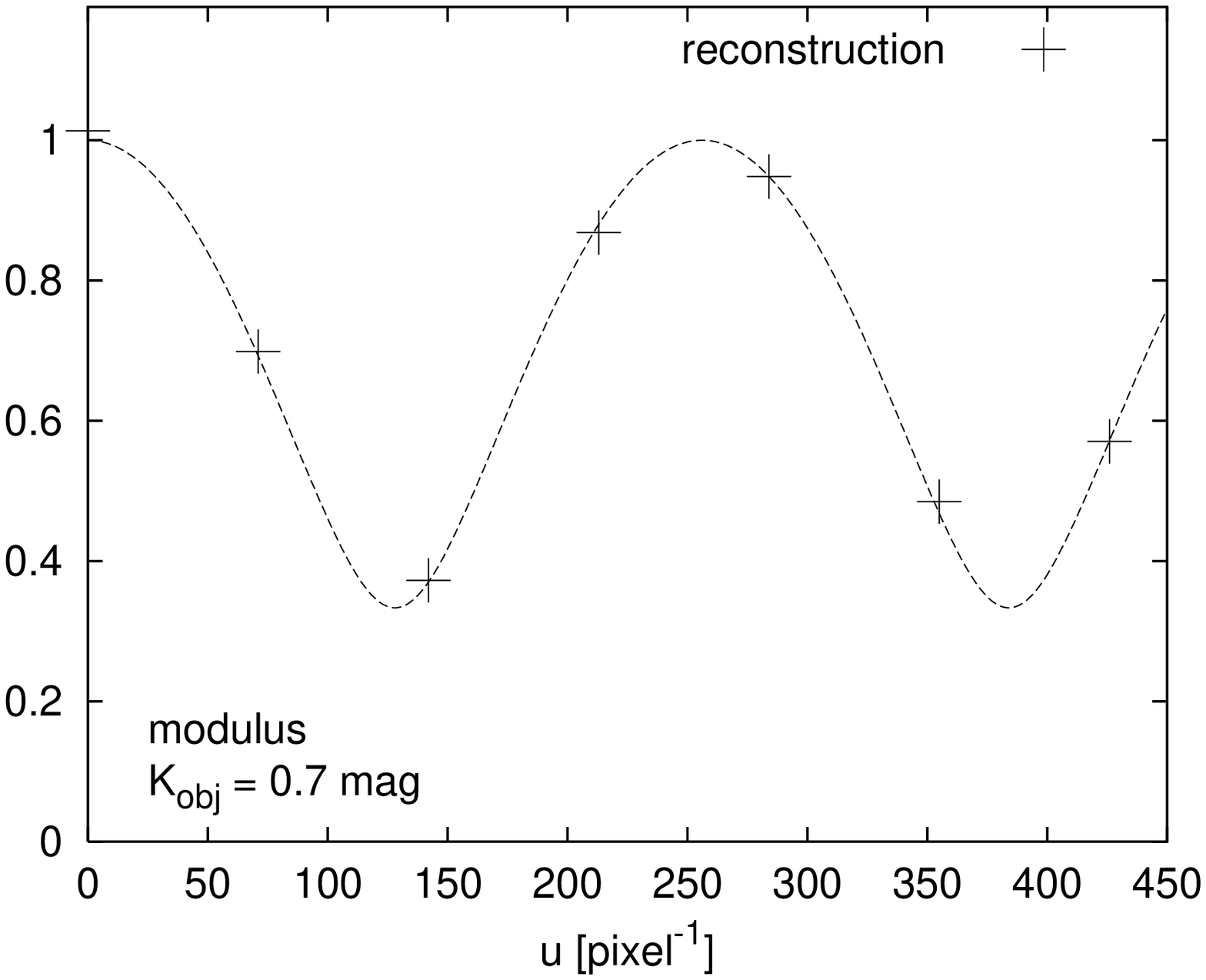,width=1.00\textwidth} \\
    \psfig{figure=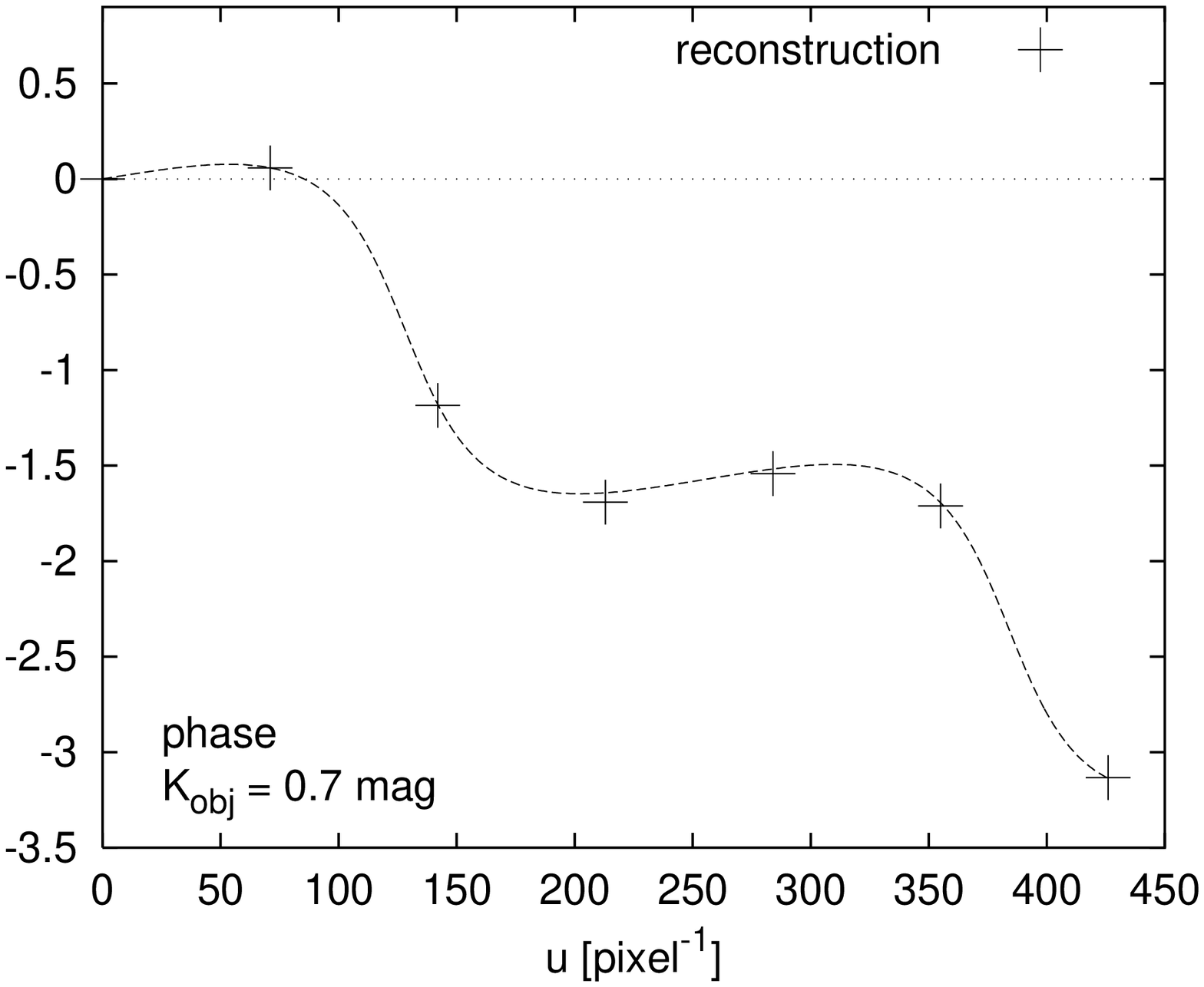,width=1.00\textwidth} \\
    \psfig{figure=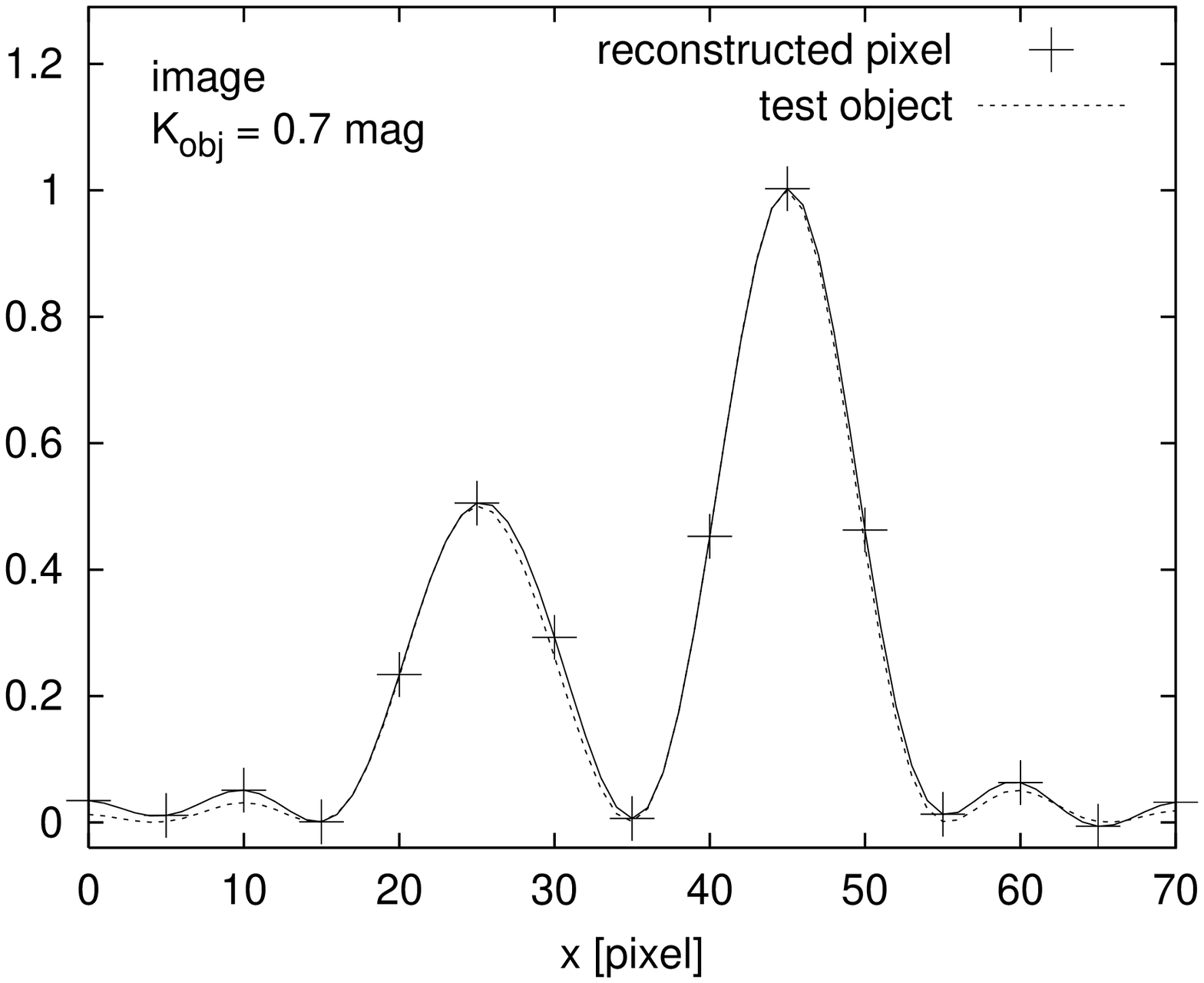,width=1.00\textwidth} \\
  \end{tabular}
\end{minipage}
\begin{minipage}{0.48\textwidth}
 \hspace{1cm} \\[3.0ex]
  \begin{tabular}{c}
    \psfig{figure=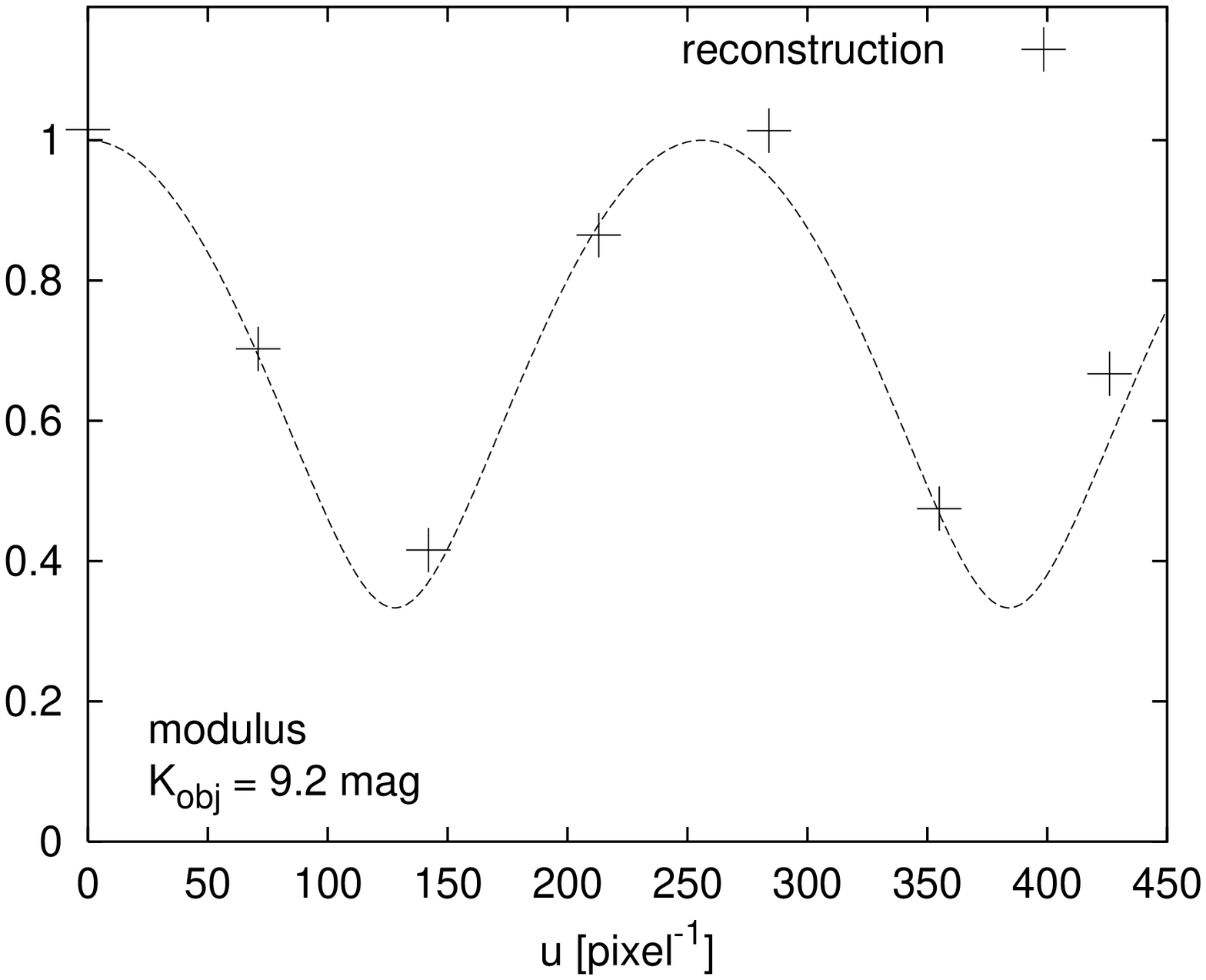,width=1.00\textwidth} \\
    \psfig{figure=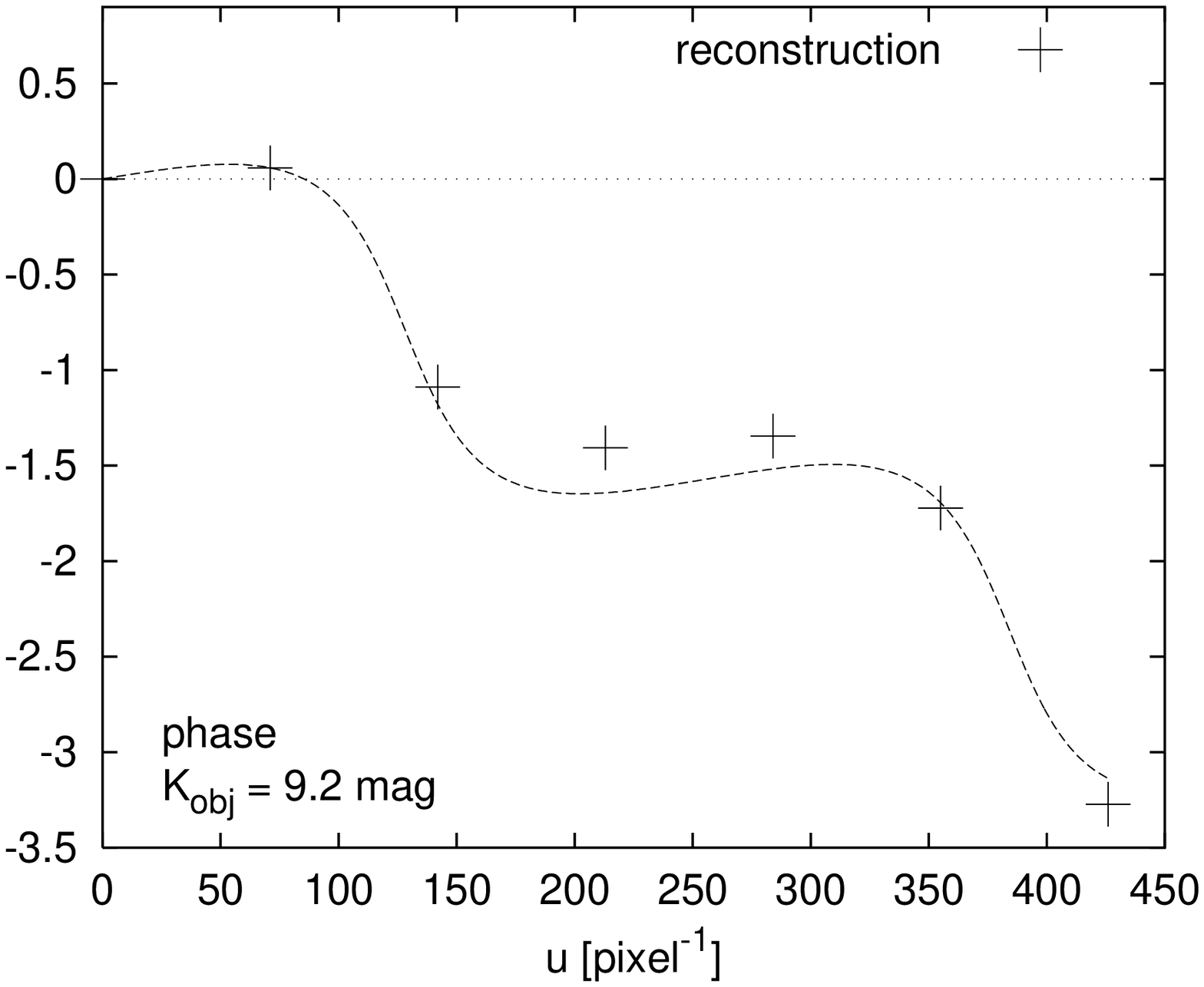,width=1.00\textwidth} \\
    \psfig{figure=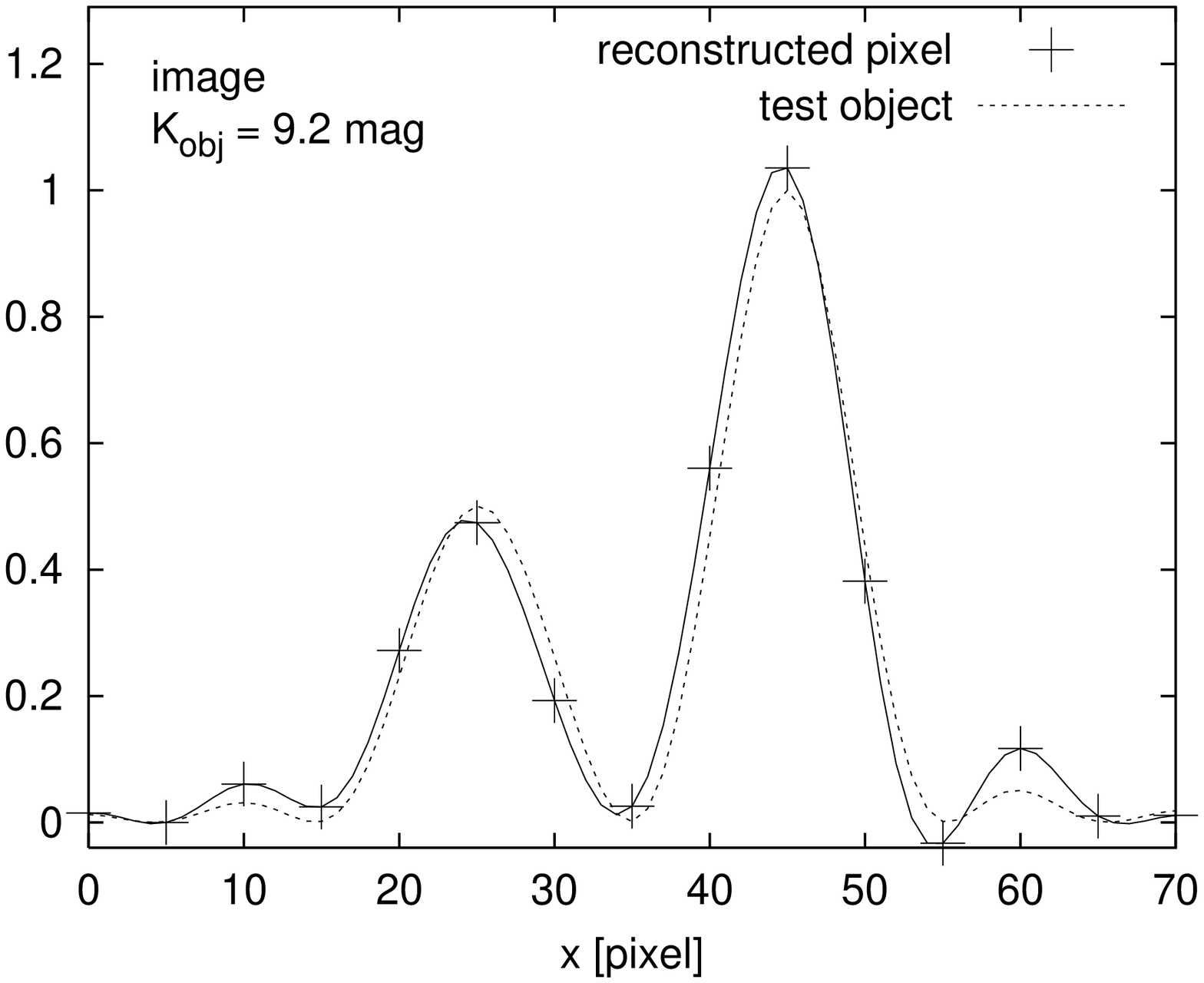,width=1.00\textwidth} \\
  \end{tabular}
\end{minipage}
\caption{\footnotesize 
Modulus ({\bf top}), phase ({\bf midlle}) and image ({\bf bottom}) of 
simulated AT-VLTI/AMBER (wide-field mode) phase-closure observations
of a binary at 2.11\,$\mu$m  if {\it three telescopes} are employed
(reconstructions are derived from interferograms of only one of the
$R/R_{K} \sim 13$ spectral channels). 
The left column refers to a total object brightness of $K=0.7$\,mag, the
right column to one of $K=9.2$\,mag.
The AT-VLTI/AMBER simulation takes 5 different telescope configurations
into account and is based on 6 statistically independent
data sets, each consisting of 1200 interferograms. 
The crosses and solid lines refer
to reconstructions based on one data set,
the dashed lines to the original object.
}
\label{Fmodphsrek}
\end{figure}

As model object 
served a binary star with an intensity distribution as shown
in Fig.~\ref{Fmodphsrek} (bottom; dashed line). The components' intensity
ratio is 1:2.
As in the case of the visibility simulations of
IRC\,+10\,420 each simulation is based on 6 statistically independent
data sets. The number of
interferograms per data set is 1200. 

\begin{figure}
\begin{minipage}{0.46\textwidth}
\begin{center}
\begin{tabular}{|c|c|c|c|c|c|} \hline 
  &   & $r_0$ [m] & $\delta_{\rm tt} [\%]$ &  $K$ [mag] & $\sigma_{p}$\\ 
\hline \hline 
       & obj. & 0.9 & 0.1   & 0.7 & \\
\rb{A'} & ref. & 0.9 & 0.1  & 0.7 & \rb{0.0054} \\ 
\hline  
       & obj. & 0.9 & 0.1   & {\bf 7.2}  & \\
\rb{K'} & ref. & 0.9 & 0.1  & 0.7 & \rb{0.0063} \\ 
\hline 
       & obj. & 0.9 & 0.1  & {\bf 8.2}  & \\
\rb{L'} & ref. & 0.9 & 0.1  & 0.7 & \rb{0.0066} \\ 
\hline 
       & obj. & 0.9 & 0.1  & {\bf 9.2}   & \\
\rb{M'} & ref. & 0.9 & 0.1  & 0.7 & \rb{0.0381} \\ 
\hline 
       & obj. & 0.9 & 0.1  & {\bf 10.2}  & \\
\rb{N'} & ref. & 0.9 & 0.1  & 0.7 & \rb{0.1580} \\ 
\hline 
\end{tabular}
\end{center}
\end{minipage}
\begin{minipage}{0.52\textwidth}
 \hspace{1cm} \\[3.0ex]
  \begin{tabular}{c}
    \psfig{figure=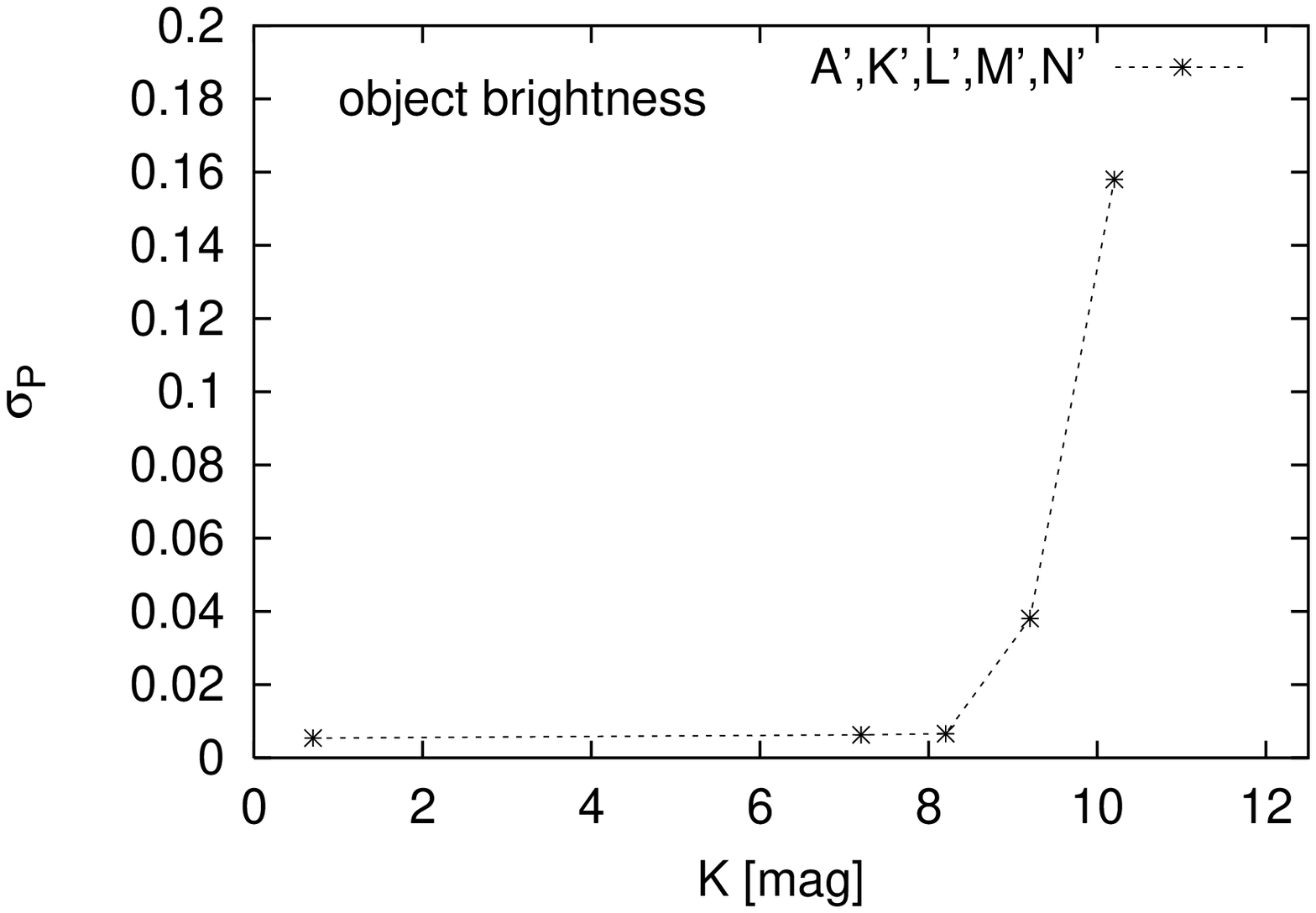,width=1.00\textwidth} \\
  \end{tabular}
\end{minipage}
\caption{\footnotesize 
Dependence of the error bars of simulated AT-VLTI/AMBER 
(wide-field mode) phase-closure observations  of a binary at 2.11\,$\mu$m on
the total object brightness if {\it three telescopes} are employed
(reconstructions are derived from interferograms of only one of the
$R/R_{K} \sim 13$ spectral channels). 
The table columns refer to the Fried paramater $r_{0}$, the residual tip-tilt
error $\delta_{\rm tt}$, the $K$-magnitude and the photometry
error $\sigma_{p}$
(i.e.\ the deviation from the components' intensity ratio of 1:2;
 based on  6 statistically independent repititions of each simulation). 
Each AT-VLTI/AMBER simulation refers to $N$=1200 interferograms
of a binary with an intensity profile as given in Fig.~\ref{Fmodphsrek}.}
\label{Fsimul3tk}
\end{figure}

Fig.~\ref{Fmodphsrek} shows the simulated
modulus, phase and image reconstruction
of the binary for total object $K$-magnitudes of 0.7\,mag and 9.2\,mag, resp.
Even for objects as faint as $K=9.2$\,mag the agreement between test object
and reconstructed object is very good. Beyond $K=10$\,mag the errors 
increase considerably. This is also illustrated in Fig.~\ref{Fsimul3tk} which
shows the photometry error
(i.e.\ the deviation from the components' intensity ratio of 1:2)
as a function of the object brightness.
The signal-to-noise ratio of the reconstruction can be increased if the
reconstructions of all $R/R_{K}$ spectral channels are averaged.


\section{Conclusions}
We have presented computer simulations of interferometric imaging with the
VLT interferometer and the AMBER instrument in the wide-field
mode. These simulations include both the
astrophysical modelling of a stellar object by radiative transfer
calculations and the simulation of  light propagation from the object to the
detector and simulation of photon noise and detector read-out noise.
We focussed on stars in late stages of stellar evolution and examplarily
studied one of its most outstanding representatives, the dusty supergiant
IRC\,+10\,420. The model intensity distribution of this key object,
obtained by radiative transfer calculations,
served as astrophysical input for the VLTI/AMBER simulations.
The results of these simulations show the dependence of the
visibility error bar on various observational parameters.
With these simulations at hand one can immediately see under
which conditions the visibility data quality would allow us to discriminate
between different model assumptions
(e.g.\ the size of the superwind amplitude $S$).
Inspection of
Fig.~\ref{FsimvisATH} shows that in all studied cases 
the observations will give clear preference to one particular model.
Therefore, observations with VLTI will certainly be well suited to gain
deeper insight into the physics of dusty supergiants.

%
%



\begin{thebibliography}{99}   

\bibitem{GlinEtal2000}
 Glindemann, A., et al., 2000, Proc. SPIE Conf. 4006-01

\bibitem{PetEtal98}
 Petrov, R., Malbet, F., Richichi, A., Hofmann, K.-H.: 1998,
   ESO Messenger 92, 11

\bibitem{PetEtal2000}
 Petrov, R., et al. , 2000, Proc. SPIE Conf. 4006-07

\bibitem{RichEtal2000}
 Richichi, A., et al., 2000, Proc. SPIE Conf. 4006-08

\bibitem{HumStrMurLow73} 
    Humphreys R.M., Strecker D.W., Murdock T.L., Low, F.J., 1973, ApJ 179, L49

\bibitem{OudGroeMatBloSah96}
    Oudmaijer R.D., Groenewegen M.A.T., Matthews H.E., Blommaert J.A.D.L,
    Sahu K.C., 1996, MNRAS 280, 1062

\bibitem{KloChePan97}
    Klochkova\,V.G., Chentsov\,E.L., Panchuk\,,V.E., 1997, MNRAS\,292,19

%
\bibitem{KnaMor85}
    Knapp G.R., Morris M., 1985, ApJ 292, 640

\bibitem{MutEtal79}
    Mutel R.L., Fix J.D., Benson J.M., Webber J.C., 1979, ApJ 228, 771

\bibitem{NedBow92}
    Nedoluha G.E., Bowers P.F., 1992, ApJ 392, 249

\bibitem{JonHumGehEtal93}
    Jones T.J., Humphreys R.M, Gehrz, R.D. et al., 1993, ApJ 411, 323

\bibitem{HumSmiDavEtal97} 
    Humphreys R.M., Smith N., Davidson K. et al., 1997, AJ 114, 2778

%
%
\bibitem{DyckEtal84}
 Dyck\,H., Zuckerman\,B., Leinert\,C., Beckwith\,S.,\,1984,\,ApJ\,287,\,801

\bibitem{RidgEtal86}
     Ridgway S.T., Joyce R.R., Connors D., Pipher J.L., Dainty C., 1986,
     ApJ 302, 662

\bibitem{CobFix87}
    Cobb M.L., Fix J.D., 1987, ApJ 315, 325

\bibitem{ChrEtal90}
Christou J.C., Ridgway S.T., Buscher D.F., Haniff C.A., McCarthy~Jr. D.W.,
  1990, Astrophysics with infrared arrays, R.\,Elston (ed.),
  ASP conf.\ series 14, p.~133

\bibitem{KastWein95}
     Kastner J., Weintraub D.A., 1995, ApJ 452, 833

\bibitem{BloeckEtal99}
 Bl\"ocker T., Balega Y., Hofmann K.-H., Lichtenth\"aler J., Osterbart R.,
 Weigelt G., 1999, A\&A 348, 805

\bibitem{Wei77}
     Weigelt G., 1977, Optics Commun. 21, 55

\bibitem{LohWeiWir83}
    Lohmann A.W., Weigelt G., Wirnitzer B., 1983, Appl. Opt. 22, 4028

\bibitem{HofWei86}
     Hofmann K.-H., Weigelt G., 1986, A\&A 167, L15

\bibitem{Lab70}
     Labeyrie A., 1970, A\&A  6, 85

\bibitem{WalEtal91}
    Walmsley C.M., Chini R., Kreysa E. et al.,
    1991, A\&A 248, 555

\bibitem{CraEtal76}
    Craine E.R., Schuster W.J., Tapia S., Vrba F.J., 1976, ApJ 205, 802

\bibitem{SavMat79}
    Savage B.D., Mathis J.S., 1979, ARA\&A 17, 73

%
%
\bibitem{MRN77}
    Mathis J.S., Rumpl W., Nordsieck K.H., 1977, ApJ 217, 425 
    
\bibitem{DraLee84}
    Draine B.T., Lee H.M., 1984, ApJ 285, 89

\bibitem{MamEtal93}
    Mampaso, A., Prieto, M. and  S\'anches, F., 1993, Infrared Astronomy,
    Cambridge University Press, p. 358

\bibitem{Rod81}
    Roddier F., 1981, Progress in Optics XIX, 281

\bibitem{Tall2}
   Tallon M., Tallon-Bosc I., 1992, A\&A 253, 641

\bibitem{MalEtal2000}
   Malbet F., et al., 2000, VLTI AMBER Instrument Analysis Review.

\bibitem{BloeckerEtal2000}
 Bl\"ocker T, Hofmann K.-H., Przygodda F., Weigelt G.,
  2000, Proc. SPIE Conf. 4006-08

\bibitem{Jen1959}
 Jennison R.C., 1959, MNRAS 118, 276

\end{thebibliography}
\end{document}